\documentclass[aps]{revtex4}
\usepackage[dvips]{graphicx}
\usepackage{amssymb,amsmath,empheq}

\RequirePackage{lineno}
\newcommand{\BB}[1]		{\mbox{\boldmath${#1}$}}

\newcommand{\Vv}		{{\BB{v}}}

\newcommand{\VD}		{{\BB{D}}}

\newcommand{\DX}		{\BB{\nabla}}

\newcommand{\Dx}		{\partial_x}
\newcommand{\Dt}		{\partial_t}

\newcommand{\mm}		{\, \rm mm}
\newcommand{\cm}		{\, \rm cm}
\newcommand{\m}			{\, \rm m}

\newcommand{\s}			{\, \rm s}

\newcommand{\ie}		{{\it i.e. }}

\newcommand{\figwidth} 	{3.0in}
\newcommand{\Dxp}		{\partial_{x^\prime}}
\newcommand{\Da}		{{\rm Da}}
\newcommand{\Pe}		{{\rm Pe}}

\newcommand*\patchAmsMathEnvironmentForLineno[1]{%
  \expandafter\let\csname old#1\expandafter\endcsname\csname #1\endcsname
  \expandafter\let\csname oldend#1\expandafter\endcsname\csname end#1\endcsname
  \renewenvironment{#1}%
     {\linenomath\csname old#1\endcsname}%
     {\csname oldend#1\endcsname\endlinenomath}}%
\newcommand*\patchBothAmsMathEnvironmentsForLineno[1]{%
  \patchAmsMathEnvironmentForLineno{#1}%
  \patchAmsMathEnvironmentForLineno{#1*}}%
\AtBeginDocument{%
\patchBothAmsMathEnvironmentsForLineno{equation}%
\patchBothAmsMathEnvironmentsForLineno{align}%
\patchBothAmsMathEnvironmentsForLineno{flalign}%
\patchBothAmsMathEnvironmentsForLineno{alignat}%
\patchBothAmsMathEnvironmentsForLineno{gather}%
\patchBothAmsMathEnvironmentsForLineno{multline}%
}

\begin{document}
\title{Instabilities in the dissolution of a porous matrix}

\author{Piotr Szymczak}
\affiliation{Institute of Theoretical Physics, Faculty of Physics, University of Warsaw,  
Ho\.{z}a 69, 00-618, Warsaw, Poland}
\email{Piotr.Szymczak@fuw.edu.pl}

\author{Anthony J.C. Ladd}
\affiliation{Chemical Engineering Department, University of Florida,
Gainesville, FL  32611-6005, USA}

\begin{abstract}

A reactive fluid dissolving the surrounding rock matrix can trigger an instability in the dissolution front, leading to spontaneous formation of pronounced channels or wormholes. Theoretical investigations of this instability have typically focused on a steadily propagating dissolution front that separates regions of high and low porosity. In this paper we show that this is not the only possible dissolutional instability in porous rocks; there is another instability that operates instantaneously on any initial porosity field, including an entirely uniform one. The relative importance of the two mechanisms depends on the ratio of the porosity increase to the initial porosity. We show that the ``inlet'' instability is likely to be important in limestone formations where the initial porosity is small and there is the possibility of a large increase in permeability. In quartz-rich sandstones, where the proportion of easily soluble material (e.g. carbonate cements) is small, the instability in the steady-state equations is dominant.

\end{abstract}

\maketitle


\section{Introduction}

Dissolution is of fundamental importance in a variety of geological systems, including diagenesis, karst formation, aquifer evolution~\citep{Ortoleva1994}, and melt migration~\citep{Aharonov1995}. It also plays an important role in a number of engineering applications, such as dam stability~\citep{Romanov2003}, ${\rm CO}_2$ sequestration~\citep{Ennis-King2007}, risk assessment of contaminant migration in groundwater~\citep{Fryar1998}, and stimulation of petroleum reservoirs~\citep{Fredd1998}. The dynamics of a dissolution front in a porous matrix  is complex, even under laminar flow conditions, with a number of possible feedback loops between reaction, diffusion and flow~\citep{Golfier2002b}. These feedback processes may trigger an instability in the front, leading to the formation of high permeability zones~\citep{Chadam1986,Sherwood1987,Hinch1990}. As a result of this {\it reactive-inflitration instability}, long, finger-like channels or ``wormholes'' are formed, which can carry active reactant deep into the matrix~\citep{Daccord1987a,Fredd1998}. 

Previous investigations of the reactive-infiltration instability~\citep{Chadam1986,Sherwood1987,Hinch1990} assumed that a steadily propagating dissolution front, separating regions of high and low porosity, forms first; subsequently an instability in this initially planar front develops. However, we have found that dissolution may be unstable from the very beginning, even before a reaction front is formed. Starting with an entirely uniform porous material, we have analyzed the competition between the development of a planar dissolution front and the growth of instabilities in that front. Here we focus on two limiting cases: first, the instability in the dissolution of an entirely homogeneous porous matrix (the ``inlet'' instability) and second, the instability in a steadily propagating reaction front (the ``front'' instability). In the convection-dominated limit the inlet instability shows a strong wavelength selection, similar to that found in dissolving fractures~\citep{Szymczak2011}, whereas the front instability is unstable over a broad range of wavelengths. However, the addition of a small diffusive flux changes the character of the front instability, introducing a strong wavelength selection in that case as well; this has been overlooked in previous work~\citep{Sherwood1987,Hinch1990}. The relative importance of the two mechanisms for unstable growth of the dissolution front depends on the amount of soluble material in the rock.

\section{Dissolution in porous media}\label{sec:eqn}

We begin with standard equations for the dissolution of a porous matrix. The  superficial fluid velocity in the pore space, $\Vv$, is given by Darcy's law:
\begin{equation}\label{eq:Darcy}
\Vv = - K(\phi)\frac{\DX p}{\mu},
\end{equation}
where the permeability $K(\phi)$ is taken as a function of the fluid fraction $\phi$ and $\mu$ is the fluid viscosity. The continuity equation takes the form
\begin{equation}\label{eq:mom}
\Dt \phi + \DX \cdot \Vv = 0,
\end{equation}
assuming the fluid filling the pore space is incompressible.

Transport of reactants and products are described by the convection-diffusion equation; for simplicity we consider a single species with a dilute concentration field $c$,
\begin{equation}\label{eq:CD}
\Dt\left(\phi c \right) + \Vv \cdot \DX c = \DX \cdot \VD(\phi, \Vv) \phi \cdot \DX c - R(c),
\end{equation}
where $\VD(\phi, \Vv)$ accounts for dispersion of reactants by flow through the porous matrix, and $R(c)$ describes the rate of consumption of reactants by the porous matrix. Finally, we have an equation for the dissolution of the porous matrix,
\begin{equation}
\label{eq:erosion}
\nu_{sol} c_{sol} \Dt \phi = \nu R(c),
\end{equation}
where $c_{sol}$ is the concentration of the solid species, and $\nu_{sol}$ and $\nu$ are stoichiometry numbers.

The boundary conditions are constructed from the following considerations. We assume that the material is initially homogeneous with a porosity $\phi_0$ and semi-infinite in extent, $0 \le x < \infty$. Far from the inlet the material is undissolved so that $v_x(x \rightarrow \infty) = v_0$, where $v_0$ is a constant; in effect we fix the far-field pressure gradient to give the desired flow velocity. A constant concentration $c_{in}$ is maintained at the inlet ($x=0$). 

We will make a number of additional simplifications which do not affect the overall conclusions, but which allow for a simpler analysis and clearer presentation. First, we assume a linear kinetic equation,
\begin{equation}
\label{eq:kin}
R(c) = -k s c,
\end{equation}
where $k$ is the reaction rate, $s$ is the specific surface area and $c$ is the reactant concentration. In this equation we assume that the reaction rate is sufficiently small that diffusion within the pore spaces can be neglected. Otherwise a more complicated constitutive equation is required to take account of mass transport within the pores. The specific surface area can be related to the grain size by assuming the grains are spherical, in which case $s = 6(1-\phi)/d$ where $d$ is the grain diameter. Although $s$ increases as the grains dissolve, to keep the derivation simple we take $s$ to be constant; then $R(c) = -k s_0 c$ with $s_0$ the specific surface area in the initial state.  We use the Carman-Kozeny equation, $K = \phi^3/5 s_0^2$, to describe the permeability, maintaining the assumption that the specific surface area remains constant during dissolution. Finally, we will only include diffusion perpendicular to the flow, with a constant diffusivity $D$. It is possible to relax any or all of these assumptions at the cost of a more involved analysis.

\section{One-dimensional analysis}\label{sec:1D}

\begin{figure}[tb]
\center\includegraphics[width=\figwidth]{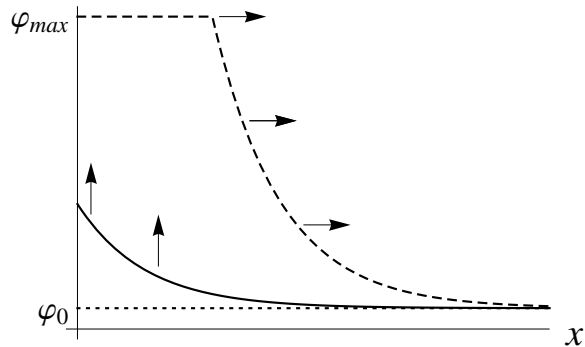}
\caption{Sketch illustrating the two regimes of dissolution: the porosity field $\phi(x)$ during the initial dissolution $t < t_{max}$, Eq.~\eqref{eq:phi1D}, is show as the solid line and the steadily advancing porosity profile $t \gg t_{max}$ \eqref{eq:phi1DS}, is shown as the dashed line.}\label{fig:instab}
\end{figure}

We first consider a one-dimensional solution of the dissolution equations~\eqref{eq:Darcy}--\eqref{eq:kin}, which will be used to establish a reference porosity profile for the linear stability analysis. Taking the flow to be in the $x$-direction, a one-dimensional porosity field evolves according to the following equations:
\begin{equation}
\label{eq:1D}
\begin{array}{c}
\Dt \phi + \Dx v_x = 0, \\
\Dt (\phi c) + v_x \Dx c = - k s_0 c, \\
\nu_{sol} c_{sol}\Dt \phi = \nu k s_0 c.
\end{array}
\end{equation}

The form of Eq.~(\ref{eq:1D}) suggests that there are characteristic length and time scales in the dissolution of the porous matrix; the penetration length of the reactant concentration $l_p = (k s_0 / v_0)^{-1}$ and the time to double the inlet porosity $t_2 = (\gamma k s_0/\phi_0)^{-1}$, where $\gamma = \nu c_{in}/(\nu_{sol} c_{sol})$. From now on we scale length by $l_p$ and time by $t_2$. In addition we scale concentration by the inlet concentration, ${\hat c} =c/c_{in}$, the superficial fluid velocity by $v_0$, ${\hat \Vv} = \Vv/v_0$, and the porosity by $\phi_0$, ${\hat \phi} = \phi/\phi_0$. Then Eq.~(\ref{eq:1D}) can be written in dimensionless form:
\begin{equation}
\label{eq:1Ds}
\begin{array}{c}
\gamma \Dt {\hat \phi} + \Dx {\hat v_x} = 0, \\
\gamma \Dt ({\hat \phi} {\hat c}) + {\hat v_x} \Dx {\hat c} = -{\hat c}, \\
\Dt {\hat \phi} = {\hat c}.
\end{array}
\end{equation}

The time scale for the evolution of the porosity is controlled by the coefficient $\gamma$, which is typically less than $10^{-2}$ in calcite dissolution, even for strong acids. Here we take the limit $\gamma \rightarrow 0$, when the velocity and concentration fields become slaved to the porosity. The superficial fluid velocity is then constant throughout the domain, ${\hat v_x}=1$, and the concentration field is time independent,
\begin{equation}
\label{eq:c1D}
{\hat c}(x) = e^{-x}.
\end{equation}
The porosity grows linearly in time,
\begin{equation}
\label{eq:phi1D}
{\hat \phi}(x,t) = 1 + t e^{-x},
\end{equation}
until only insoluble material remains; \ie when ${\hat \phi}(x,t) = \phi_{max}$, the maximum porosity (relative to its initial value).

Equation~\ref{eq:phi1D} remains valid for (dimensionless) times up to 
\begin{equation}\label{eq:t1D}
t_{max} = \phi_{max} - 1 = \Delta,
\end{equation}
at which point the reaction at the inlet stops. For times larger than $t_{max}$ a propagating front develops, which moves through the porous matrix. Although the full solution is complicated, even in one-dimension, we can consider the long-time limit where a front propagates steadily  with a constant velocity $v_f$. In this case we  create a new coordinate system relative to the front position $x_f=v_f t$, which is defined as the rightmost point where ${\hat \phi} = \phi_{max}$,
\begin{equation}
\label{eq:xf}
x^\prime(x,t) = x - v_f t.
\end{equation}
The dimensionless dissolution equations in the new coordinate system are (still in the limit $\gamma \rightarrow 0$):
\begin{equation}
\label{eq:1DsS}
\begin{array}{c}
\Dxp {\hat v}_{x^\prime} = 0, \\
{\hat v}_{x^\prime} \Dxp \hat{c} = -\hat{c}, \\
\Dt {\hat \phi} -v_f \Dxp {\hat \phi} = \hat{c}.
\end{array}
\end{equation}
At steady-state ($\Dt {\hat \phi}=0$) the porosity profile is again exponential,
\begin{equation}
\label{eq:phi1DS}
{\hat \phi}(x^\prime) = 1 + {v_f}^{-1} e^{-x^\prime}.
\end{equation}
The boundary condition at the front ${\hat \phi}(x'=0) = \phi_{max}$ determines the velocity of the front,
\begin{equation}\label{eq:uf}
 v_f = {\Delta}^{-1},
\end{equation}

A steadily advancing porosity profile, Eq.~\eqref{eq:phi1DS}, is sketched in Fig.~\ref{fig:instab} alongside the initial ($t < t_{max}$) profile, Eq.~\eqref{eq:phi1D}. Surprisingly, both profiles are unstable with respect to  infinitesimal perturbations, although the nature of the instabilities is qualitatively different in the two cases.

In our analysis we neglect any coupling between the development of the dissolution front and mechanical compaction. In a related problem of the channeling instability in an upwelling mantle~\citep{Aharonov1995}, mechanical compaction was shown to stabilize long-wavelength perturbations without affecting the value of the most unstable wavelength.

\section{Linear stability analysis}\label{sec:lsa}

We develop a linear stability analysis by considering a small sinusoidal perturbation about the one-dimensional solution;
\begin{equation}
{\hat \phi}(x,y,t) = {\hat \phi}_r(x,t) + g(x) \sin(uy) e^{\sigma t}
\end{equation}
where ${\hat \phi}_r$ is the one-dimensional reference porosity field given by Eq.~\eqref{eq:phi1D} or Eq.~\eqref{eq:phi1DS} and $g(x)$ is the amplitude of the perturbation; the dimensionless wavevector $u = 2 \pi /\lambda$, where $\lambda$ is the wavelength (in units of $l_p$). Note that the base state of the inlet instability, Eq.~\eqref{eq:phi1D}, is itself time-dependent; here we use an approximation~\citep{Tan1986} in which the base state is frozen at a specific time, $t_0$, and the growth rate is then determined as if the base state were time-independent (the quasi-steady-state approximation). A solution of the inlet instability problem for the analogous case of a narrow fracture can be found in~\citet{Szymczak2011}. In a similar fashion, after some straightforward but lengthy manipulations, we obtain a third order differential equation for the perturbation $g$,
\begin{equation}
\label{eq:dphi}
\left(u^2 - \Dx^2 + \frac{3 {\hat \phi}_r^\prime}{{\hat \phi}_r}\Dx\right)\left(\Dx + H u^2 {\hat \phi}_r\right)e^x \sigma g = \frac{3 u^2}{{\hat \phi}_r} g,
\end{equation}
where ${\hat \phi}_r^\prime = \Dx {\hat \phi}_r$. The dimensionless constant
\begin{equation}
\label{eq:H}
H = {D k s_0 \phi_0}/{v_0^2}
\end{equation}
can be written as the ratio of the Damk\"ohler number $\Da = k/v_0$ and the P\'eclet number $\Pe = v_0/D s_0 \phi_0$, $H = \Da \Pe^{-1}$.

\begin{figure}[t]
\includegraphics[width=3in]{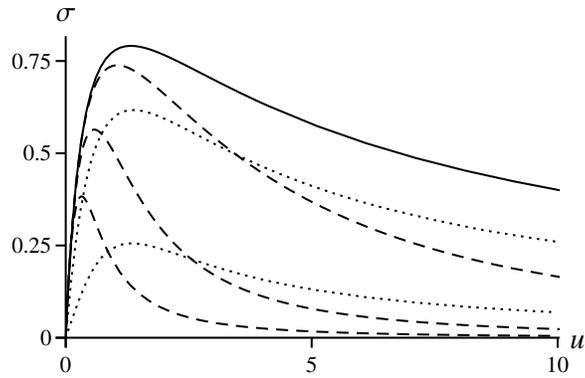}
\caption{Growth rates of the inlet instability. The solid line is for the reference state frozen at $t_0 = 0$ without diffusion ($H = 0$). The dashed lines show results for increasing $H$: top to bottom $H = 0.1$, $H = 1$, and $H = 10$. The dotted lines  show results for the reference state frozen at later times: top to bottom $t_0 = 1$ and $t_0 = 10$ with $H = 0$.}\label{fig:inlet}
\end{figure}

Boundary conditions on $g$ follow from the boundary conditions on the porosity, pressure, and concentration fields:
\begin{equation}
\label{eq:BCs}
\begin{array}{c}
g(x=0) = 0, \\
\left[\Dx^2 e^x g(x)\right]_{x=0} = 0, \\
\left[\Dx e^x g(x)\right]_{x\rightarrow\infty} = 0,
\end{array}
\end{equation}
There are three independent solutions for $g(x)$ but one of these is eliminated by the far-field boundary condition. Since the amplitude of $g$ is arbitrary only one more condition is needed to fix the solution. The final boundary condition is then used to obtain the dispersion relation for the growth rate $\sigma(u)$. Results for the growth rate of the inlet instability are shown in Fig.~\ref{fig:inlet} for different values of $H$ and $t_0$.

\begin{figure}[t]
\includegraphics[width=\figwidth]{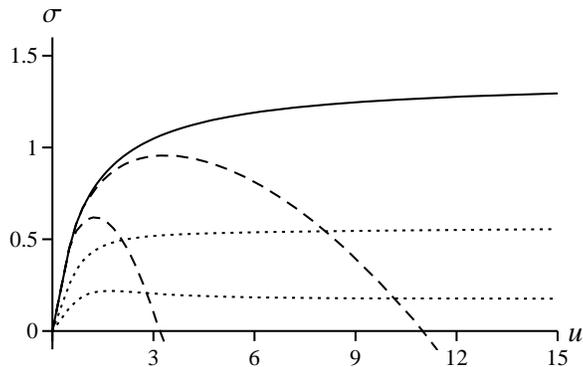}
\caption{Growth rates of the front instability. The solid line shows the case $\Delta \rightarrow 0$, without diffusion ($H=0$). The dashed lines indicate increasing diffusion: top to bottom $H = 0.001$, and $H = 0.01$. The dotted lines show the effect of increasing $\Delta$: top to bottom $\Delta = 2$ and $\Delta = 10$ with $H = 0$.}\label{fig:moving}
\end{figure}

The reference porosity field for a steadily-moving front, Eq.~\eqref{eq:phi1DS}, is time independent in the coordinate system relative to the front. The resulting equation for the perturbation in porosity is fourth order,
\begin{equation}
\label{eq:dphiS}
\left(\!u^2 - \Dx^2 + \frac{3 {\hat \phi}_r^\prime}{{\hat \phi}_r}\Dx\right) \left(\Dx + H u^2 {\hat \phi}_r\right)\!e^x \!(\sigma - v_f\Dx) g = \frac{3 u^2}{{\hat \phi}_r}g.
\end{equation}
In this case there are four independent solutions for $g(x)$, but two of these are always eliminated by the far-field boundary condition. Care must be exercised in applying the remaining two boundary conditions at the front itself, which is no longer planar but has small sinusoidal perturbations as well. As in the inlet instability, one boundary condition is sufficient to fix the solution to within a multiplicative constant, and the final boundary condition then provides the dispersion relation illustrated in Fig.~\ref{fig:moving}.

\section{Discussion}

The solid lines in Figs.~\ref{fig:inlet} and~\ref{fig:moving} indicate the most rapid growth of each instability: the inlet instability at the onset of dissolution ($t_0 \rightarrow 0$) and the steady-state instability in the limit that the front develops instantaneously ($\Delta \rightarrow 0$). In both cases we are considering the convective limit $H = 0$. The inlet instability shows a strong wavelength selection, with a maximum growth rate at a wavenumber $u_{max} \approx 1.3$. Even when the flow rate is small ($H > 1$) the growth rate is similar to the convection-dominated case, although the peak is pushed to longer wavelengths. On the other hand the front instability grows across a broad spectrum of wavelengths, limited only by the eventual onset of diffusional stabilization (dashed lines). Diffusion has a much larger effect on the front instability than on the inlet instability. Even for weak diffusion, $H=0.001$, the shape of the dispersion curve changes qualitatively; short wavelengths are stabilized and a maximum growth rate appears. This suggests that the diffusionless limit considered in earlier work~\citep{Hinch1990,Sherwood1987} is singular, and any amount of diffusion will lead to the appearance of a maximum in the dispersion curves.

The solid curve in Fig.~\ref{fig:inlet} was obtained by freezing the base state at $t_0=0$. If the base state is frozen at a later time, then the growth rate is reduced as shown by the dotted curves. However, the peak growth rate remains at almost the same wavelength, independent of $t_0$, so a single mode ($u \approx u_{max}$) dominates until the onset of non-linear growth. In the case of the front instability, the growth rate is sensitive to the porosity contrast across the dissolution front (Fig.~\ref{fig:moving}). For small $\Delta$, the growth rate increases monotonically towards a limiting value, in agreement with~\citet{Hinch1990}. However, as $\Delta$ increases, the dispersion curve flattens, and at around $\Delta \approx 3.2$ a maximum appears even in the convective limit, indicating a similar competition between reaction and convection as in the inlet instability. Interestingly, the most unstable wavelength in the front instability approaches that for the inlet instability when $\Delta \gg 1$.

The analysis outlined in Sec.~\ref{sec:lsa} includes two parameters, the ratio of Damk\"ohler and P\'eclet numbers, $H = \Da \Pe^{-1}$, and the porosity contrast, $\Delta = \phi_{max} -1$. Fracture dissolution, where an initially small (less than $1 \mm$) aperture opens essentially without limit, can be considered as the limiting case $\Delta \rightarrow \infty$. In this case a steadily advancing front never develops and Eq.~\eqref{eq:dphi} applies. We can obtain the corresponding equations for fracture dissolution by mapping the porosity ${\hat \phi}$ onto the dimensionless fracture aperture ${\hat h} = h/h_0$; the initial fracture aperture $h_0$ is related to the specific surface area by $h_0 = 2/s_0$. For a typical calcite fracture -- reaction rate $k = 2.5 \times 10^{-5} \cm s^{-1}$, aperture $h_0 = 0.2 \mm$ and hydraulic gradient of $1\%$~\citep{Dreybrodt1996} -- the dissolution is convection dominated ($H \sim 10^{-5}$) and the growth rate of the instability follows the solid line in Fig.~\ref{fig:inlet}. Thus, for these parameters we would expect to see fingers with an initial spacing of the order of $2 \pi l_p/u_{max} \approx 1 \m$, but the competition for flow between growing wormholes will eliminate many small channels and the spacing observed at later times will be larger. We have previously  suggested that this instability in the fracture dissolution front initiates the formation of underground caves in limestone~\citep{Szymczak2011}.

A convection-dominated instability can also develop in highly porous and permeable limestone formations. Crinoidal limestone, for example, can have a specific surface area as low as $100 \cm^{-1}$ and a permeability as high as $10^{-8} \cm^2$ \citep{Noiriel2004}. For a gravitationally-driven flow, $H \sim 10^{-3}$ and the penetration length is of the order of $1 \cm$. The corresponding wavelengths of an unstable front are of the order of several centimeters for the inlet instability and one or two centimeters for the front instability. The inlet instability is active from the beginning of the dissolution process with fingers developing on a dimensionless time scale $t \sim 1$. At the same time there is uniform dissolution as indicated by Eq.~\eqref{eq:phi1D}, which in the absence of the inlet instability would establish a dissolution front on a dimensionless time scale $t \sim \Delta$, Eq.~\eqref{eq:t1D}. The relative importance of the two instabilities will therefore depend on the porosity contrast $\Delta$, which typically ranges from $1$ ($\phi_0 = 0.5$) to $10$ ($\phi_0 = 0.09$).  For large values of $\Delta$, we would expect substantial fingering from the inlet, but in more porous formations a steadily advancing front would appear first.

At the opposite extreme from fracture dissolution is the limit $\Delta \rightarrow 0$ where the porosity contrast is very small, although there can still be a substantial change in permeability. This limit includes sandstones, where the inert quartz grains are interspersed with a soluble cement~\citep{Lund1976,Lund1976a}. In such cases a steadily advancing front will always form prior to the instability.

For smaller grain sizes ($s_0 \sim 10^4 \cm^{-1}$) the flow rate is extremely small ($H \gg 1$) and a depletion layer develops {\em behind} a steadily propagating front~\citep{Chadam1986}. The concentration field in the region behind the front is significantly perturbed by diffusion of solute from the reaction zone and we we can no longer assume a constant concentration at the front. In this diffusive limit a uniform front is stable at low flow rates~\citep{Chadam1986}, but above a critical flow rate, fingers are predicted to develop with a wavelength of order $D/v_0$.

Experimental observations of dissolution patterns in porous media~\citep{Daccord1987a,Fredd1998,Golfier2002b} are typically made under conditions corresponding to reservoir acidization, so the reaction rates are much higher than is typical under geophysical conditions. For example, when acidizing calcite with hydrochloric acid the reaction rate is of the order of $0.1 \cm \s^{-1}$~\citep{Fredd1998}. Taking $s_0 = 10^4 \cm^{-1}$, corresponding to micron-sized grains, and $\gamma = 0.01$, the front is established on a time scale of $(k s_0 \gamma)^{-1} \approx 0.1 \s$. Similar considerations apply to other experiments and indeed to acidization in the field. In acidization it seems clear that the instability develops out of a steadily propagating front~\citep{Hinch1990,Sherwood1987}. We reiterate that diffusion cannot be ignored in the acidization instability, even under high flow rate conditions (Fig.~\ref{fig:moving}).

\begin{acknowledgments}
This work was supported by the US Department of Energy, Chemical Sciences, Geosciences and Biosciences Division, Office of Basic Energy Sciences (DE-FG02-98ER14853).
\end{acknowledgments}



\end{document}